\documentclass[conference]{IEEEtran}
\IEEEoverridecommandlockouts

\usepackage{cite}
\usepackage{amsmath,amssymb,amsfonts}
\usepackage{algorithmic}
\usepackage{graphicx}
\usepackage{textcomp}
\usepackage{xcolor}
\usepackage{multirow}
\usepackage{booktabs}
\usepackage{caption}
\usepackage[a4paper,top=0.7in,bottom=1in,left=1in,right=1in]{geometry}

\usepackage{tikz}
\usepackage{pgfplots}
\pgfplotsset{compat=1.18} 

\usepackage[caption=false,font=footnotesize]{subfig} 
\usepackage{graphicx}
\usepackage{graphicx, subcaption}

\usepackage[caption=false,font=footnotesize]{subfig}

\usepackage{mathtools} 

\usepackage[hyphens]{url} 
\usepackage{tikz}
\usetikzlibrary{positioning,arrows.meta,calc,fit}
\usetikzlibrary{shapes.geometric,shapes.symbols} 
\usepackage{booktabs,siunitx,tabularx} 
\makeatletter
\def\IEEEbibitemsep{0pt plus 1pt}
\makeatother

\newcommand{\RightColNudge}{\par\setlength{\leftskip}{2em}}

\begin{document}


\title{
Dictionary-Based Contrastive Learning for GNSS Jamming Detection
}

\author{
\IEEEauthorblockN{
Zawar Hussain\IEEEauthorrefmark{1}\IEEEauthorrefmark{2}\textsuperscript{\dag},
Arslan Majal\IEEEauthorrefmark{3}\textsuperscript{\dag},
Aamir Hussain Chughtai\IEEEauthorrefmark{4},
Talha Nadeem\IEEEauthorrefmark{1}
}
\IEEEauthorblockA{
\IEEEauthorrefmark{1}Department of Electrical Engineering, Lahore University of Management Sciences (LUMS), Pakistan \\
\IEEEauthorrefmark{2}School of Engineering, University of Management and Technology (UMT), Pakistan \\
\IEEEauthorrefmark{3}Department of Mechanical Engineering, University of Wisconsin–Madison, USA \\
\IEEEauthorrefmark{4}Institute of Data Science, University of Engineering and Technology (UET), Pakistan \\
Emails: zawar.hussain@lums.edu.pk,\ majal@wisc.edu,\ chughtaiah@uet.edu.pk,\ talha.nadeem@lums.edu.pk
}
\thanks{\textsuperscript{\dag}Authors contributed equally.}
}
\maketitle

\begin{abstract}
Global Navigation Satellite System (GNSS) signals are fundamental in applications across navigation, transportation, and industrial networks. However, their extremely low received power makes them highly vulnerable to radio-frequency interference (RFI) and intentional jamming. Modern data-driven methods offer powerful representational power for such applications, however real-time and reliable jamming detection on resource-limited embedded receivers remains a key challenge due to the high computational and memory demands of the conventional learning paradigm. To address these challenges, this work presents a dictionary-based contrastive learning (DBCL) framework for GNSS jamming detection that integrates transfer learning, contrastive representation learning, and model compression techniques. The framework combines tuned contrastive and dictionary-based loss functions to enhance feature separability under low-data conditions and applies structured pruning and knowledge distillation to reduce model complexity while maintaining high accuracy. Extensive evaluation across varying data regimes demonstrate that the proposed algorithm consistently outperforms modern CNN, MobileViT, and ResNet-18 architectures. The framework achieves a substantial reduction in memory footprint and inference latency, confirming its suitability for real-time, low-power GNSS interference detection on embedded platforms.
\end{abstract}


\section{Introduction}

Global Navigation Satellite System (GNSS) services enable critical applications such as air navigation, autonomous driving, financial timing, and power grid control \cite{Zidan2020GNSS}. Despite their ubiquity, GNSS received signals are very weak, which makes them highly vulnerable to radio-frequency interference (RFI) and deliberate jamming \cite{10701871}. Even low-power interference can distort the correlation process or disrupt timing accuracy, potentially leading to operational failures in dependent infrastructures \cite{Savolainen2024Towards}. Consequently, there is an increasing demand for real-time, low-cost, and robust jamming detection solutions suitable for embedded platforms \cite{GNSS_Edge}.The growing sophistication of jamming attacks has exposed the inability of traditional GNSS interference mitigation and detection techniques in adapting to dynamic environments, prompting the shift toward data-driven and machine-learning-based solutions \cite{Rados2024Recent}. 

Machine/deep learning (ML/DL) techniques can be effectively leveraged for GNSS interference detection, as they can learn complex, non-linear signal characteristics and adapt to dynamic interference conditions more efficiently than traditional rule-based or model-driven approaches \cite{Mehr2025A, Heublein2024Evaluating, Ivanov2024Deep}. Deep learning architectures that operate directly on spectrograms such as Hybrid CNN-Transformer models can effectively model the spectral–temporal structure of GNSS signal, enabling robust classification performance in complex interference environments \cite{article}. Conventional learning models require substantial computational, memory, and energy resources, limiting their deployment on edge platforms such as IoT devices and low-power receivers \cite{Shuvo2023Efficient}. To enable effective use in such settings, models must be optimized for compactness, fast inference, and efficient generalization.

We can use off-the-shelf DL methods customized for edge devices. MobileViT and its variants are viable lightweight go-to options in this regard, which also have significant potential for real-time jamming detection  \cite{mobie_vit}. Further, we can leverage recent advances in model compression techniques, such as structured pruning \cite{he2017channel}, quantization, and knowledge distillation (KD) \cite{hinton2015distilling}, to reduce model complexity of classical DL methods with large memory overhead while preserving critical representational power.

Recent advancements in the field of contrastive learning has highlighted its potential to significantly enhance feature learning, particularly in low-data regimes. Contrastive frameworks such as SimCLR\cite{chen2020}, Supervised Contrastive (SupCon) \cite{khosla2020supervised}, Tuned Contrastive Learning (TCL) \cite{animesh2025tuned}, and Cross-Entropy with Contrastive Fusion (CLCE) \cite{long2024} learn semantically structured embeddings by maximizing inter-class separation and minimizing intra-class variance.

Building on the strength of contrastive learning, we propose a Dictionary-Based Contrastive Learning (DBCL) framework tailored for GNSS jamming detection. Integrating transfer learning, contrastive embedding, and model compression techniques such as pruning and distillation to achieve model compactness and efficiency. \cite{xu2024attention}. We conduct extensive simulations using the proposed DBCL model, which consistently outperforms conventional CNN, ResNet-18, MobileViT, and hybrid baselines across both low- and high-sample training conditions. Achieving a substantial memory footprint reduction (approximately 75\%) and inference latency while maintaining high classification accuracy, confirming its suitability for real-time, energy-efficient GNSS jamming detection on embedded edge platforms. 


We contribute in the following ways: (1) We develop a transfer-learning-based GNSS interference classification framework utilizing dictionary-based contrastive embeddings for robust representation learning. (2) We incorporate model compression through pruning and distillation for deployment on edge/embedded devices. (3) We  demonstrate significant improvements in accuracy, generalization, and efficiency across small data regimes. (4) We validate model’s real-time feasibility for edge-based GNSS jamming detection.

The rest of this paper is organized as follows: Sections II and III comprise data preprocessing and the proposed methodology. Section IV presents the simulation results followed by conclusion in section V.

\section{Data Preprocessing}

To validate the proposed framework, we employ the publicly available GNSS interference dataset from \cite{ghanbarzade2025gnss}, which includes complex baseband (IQ) recordings of clean and jammed signals under diverse interference scenarios. Before training, all recordings are transformed into a 3-channel Short-Time Fourier Transform (STFT) image corpus consisting of log-magnitude, cosine phase, and sine phase representations for training, validation, and testing. For each class, recordings within the designated training and test are enumerated, ensuring class consistency across splits. We define per-class \emph{targets} specifying the number of fixed-length segments for each split, reserving a validation subset (200 recordings by default) before segmentation. Each channel '$v$' of in-phase and quadrature sequences \(I, Q\in\mathbb{R}^{N}\) is independently standardized (z-score): \(v \leftarrow (v-\mu_v)/(\sigma_v+\varepsilon)\) with \(\varepsilon=10^{-6}\). Segments are generated by sliding a window of length \(L=\lfloor \texttt{crop\_sec}\times \texttt{sr}\rfloor\) samples with hop size \(H=\max\{1,\lfloor L(1-\texttt{overlap})\rfloor\}\), yielding either non-overlapping or partially overlapping crops depending on the overlap fraction. To meet per-class targets, \emph{crop quotas} are computed deterministically for each recording, ensuring balanced class distribution. If a recording yields fewer distinct windows than its quota, additional segments are generated by reproducible circular shifts of the standardized \(I/Q\) streams, ensuring no inter-class imbalance.

Each crop is processed using STFT with a Hann window to generate the complex spectrogram \( \mathcal{S}=\mathcal{S}_I+j\mathcal{S}_Q\), where \(\mathcal{S}_I=\mathrm{STFT}(I)\) and \(\mathcal{S}_Q=\mathrm{STFT}(Q)\). The magnitude \(|\mathcal{S}|\) and phase \(\angle\mathcal{S}\) are computed for each time–frequency bin. The three output channels are
$
\text{ch}_0=\log\!\big(1+|\mathcal{S}|\big),\quad
\text{ch}_1=\cos(\angle\mathcal{S}),\quad
\text{ch}_2=\sin(\angle\mathcal{S}).
$
Magnitude is mapped to \([0,1]\) by percentile scaling (0.5–99.5\,\%), and phase-derived channels are affinely mapped from \([-1,1]\) to \([0,1]\). These channels are stacked, converted to 8-bit PNG, and resized to \(\texttt{img\_size}\times \texttt{img\_size}\). The resulting images are stored in class-named directories for training, validation, and testing. The preprocessing parameters are summarized in Table~\ref{tab:prep-params}, and all operations such as file iteration, crop selection, and STFT construction are performed with fixed seeds to ensure deterministic results.

\begin{table}[t]
\centering
\caption{Preprocessing and STFT parameters}
\label{tab:prep-params}
\resizebox{\columnwidth}{!}{%
\begin{tabular}{ll}
\toprule
\textbf{Parameter} & \textbf{Value / Definition} \\
\midrule
Sample rate (\texttt{sr}) & 16{,}000 Hz \\
Crop duration (\texttt{crop\_sec}) & 1.0 s \\
Overlap fraction (\texttt{overlap}) & 0.5 $\Rightarrow$ hop $H=\lfloor L(1-0.5)\rfloor$ \\
Crops per file (train / val / test) & 4 / 5 / 4 \\
Window / FFT size & Hann, $\texttt{n\_fft}=1024$ \\
Hop size (STFT) & hop = 256 samples \\
Complex spectrogram & $\mathcal{S}=\mathcal{S}_I + j\,\mathcal{S}_Q$ \\
Channel mapping & $\log(1+|\mathcal{S}|),~\cos\angle\mathcal{S},~\sin\angle\mathcal{S}$ \\
Magnitude scaling & Percentile norm.\ (0.5–99.5\%) to [0,1] \\
Phase scaling & Linear map from [-1,1] to [0,1] \\
Image size (\texttt{img\_size}) & 224$\times$224 (bilinear) \\
Standardization & Per-stream z-score on $I$ and $Q$ \\
Validation hold-out & 200 files/class (by file, before cropping) \\
\bottomrule
\end{tabular}%
}
\end{table}

\begin{figure*}[t]
  \centering
  \includegraphics[width=\textwidth]{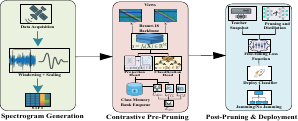}
  \caption{Proposed DBCL Architecture}
  \label{fig:prep-model}
\end{figure*}

\section{Proposed Methodology}

This section provides an overview of our proposed framework, followed by spectrum generation, pre-pruning, and post-pruning deployment details.  

Given a three–channel time–frequency tensor \(\mathbf{X}\in\mathbb{R}^{3\times H\times W}\) (magnitude, \(\cos\phi\), \(\sin\phi\)), our model maps \(\mathbf{X}\) to a class label via a convolutional backbone and a linear classifier, and simultaneously learns a normalized embedding through supervised contrastive learning. The final training objective couples a smoothed cross-entropy term with a contrastive term; during model compression, we apply structured channel pruning followed by knowledge distillation (KD)  from a frozen pre-pruning teacher.

\subsection{Spectrogram Generation}

In the first stage of our proposed framework as shown in Fig.~\ref{fig:prep-model}, the spectrogram generation process begins with data acquisition, where raw GNSS signal is captured as complex IQ sequences. After windowing and scaling, the signal is then processed through STFT, producing a spectrogram. This spectrogram serves as the input for our model, where it is further processed to extract magnitude (log) and phase (cosine and sine), forming the three output channels used by our model.

\subsection{Pre-Pruning Stage} The pre-pruning stage focuses on learning discriminative and compact feature representations through contrastive pre-training using a ResNet-18 backbone and a dual-headed architecture (classification and projection).

\subsubsection{\textbf{Backbone and Heads}}
A ResNet-18 backbone $f_\theta(\cdot)$ \cite{he2016deep} extracts spatial–spectral features from each input spectrogram, producing a feature vector $\mathbf{h}=f_\theta(\mathbf{X})\in\mathbb{R}^{512}$. 
Two heads are attached:
\begin{itemize}
    \item \textbf{Classification Head:} The classification head takes the feature vector \( \mathbf{h} \) produced by the ResNet-18 backbone and maps it to class logits \( \mathbf{o}_s^{(v)} \in \mathbb{R}^C \), where \( C \) is the number of classes. Given an input time-frequency image \( \mathbf{X} \) and its corresponding ground-truth label \( y \in \mathcal{C} \), the network computes the logits: \(\mathbf{o}_s^{(v)} = c(f_{\theta}(\mathbf{X}^{(v)})) \in \mathbb{R}^{C}\) and the predictive distribution: \(p_{\theta}(\mathbf{y}\mid\mathbf{X}^{(v)})=\mathrm{softmax}\!\big(\mathbf{o}_s^{(v)}\big)\). For each stochastic view \( \mathbf{X}^{(v)} \) (two views are used, \( \mathcal{V}_{\mathrm{cls}} = \{1, 2\} \)), we compute the \emph{cross-entropy loss} with \emph{label smoothing}. Label smoothing with parameter \( \varepsilon \) \cite{hinton2015distilling} assigns probability \( 1 - \varepsilon \) to the true class \( y \), and distributes \( \varepsilon \) uniformly over the remaining \( C-1 \) classes. The view-averaged cross-entropy loss is given by:
\begin{align}
\mathcal{L}_{\mathrm{CE}}
&= \frac{1}{|\mathcal{V}_{\mathrm{cls}}|}\!
\sum_{v\in\mathcal{V}_{\mathrm{cls}}}
\Big[-\, q_{\varepsilon}(\cdot\mid y)^{\top}\log p_{\theta}\big(\mathbf{y}\mid\mathbf{X}^{(v)}\big)\Big].
\end{align}
This loss encourages the model to output probabilities that match the smoothed target, improving generalization by reducing overconfidence in the model’s predictions.
    \item \textbf{Projection Head:} Two linear layers with ReLU and normalization map $\mathbf{h}$ to a normalized embedding $\mathbf{z}=\mathrm{norm}(g(\mathbf{h}))\in\mathbb{R}^{d}$ used only for contrastive training. This head is discarded after training without affecting inference accuracy.
\end{itemize}

\subsubsection{\textbf{Two-View Contrastive Training}}
Each training sample is augmented into two stochastic views, $(\mathbf{X}^{(1)},\mathbf{X}^{(2)})$, using small time shifts, frequency masking, and mild magnitude jitter. Both views share the same ground-truth label, encouraging invariance to local distortions in the time–frequency domain.

\subsubsection{\textbf{Tuned Contrastive Learning (TCL)}}

TCL enhances supervised contrastive learning by adaptively weighting hard positives and hard negatives to refine class-wise separability. 
Each training sample is presented as two stochastic views, 
$\mathcal{V}_{\mathrm{cls}}=\{1,2\}$, with $\ell_2$-normalized embeddings 
$\mathbf{z}^{(v)}\in\mathbb{R}^{d}$ obtained from the projection head.
Let $\mathcal{B}$ denote the index set of all view-level embeddings in the mini-batch. 
For an anchor $i\in\mathcal{B}$ with label $y_i$, positives and negatives are defined as
\begin{align}
\mathcal{P}(i) &= \{\,j\in\mathcal{B}\setminus\{i\}~|~ y_j = y_i \,\}, \\
\mathcal{N}(i) &= \{\,j\in\mathcal{B}\setminus\{i\}~|~ y_j \neq y_i \,\}.
\end{align}
Cosine similarity between embeddings is computed as
\begin{align}
s_{ij} = \frac{\mathbf{z}_i^\top \mathbf{z}_j}{\tau},
\end{align}
where $\tau>0$ is a temperature parameter. TCL modifies the contrastive denominator with tunable coefficients 
$k_1,k_2\ge0$ to emphasize hard positive and hard negative pairs:
\begin{align}
D_i = 
\underbrace{\sum_{p\in\mathcal{P}(i)} e^{s_{ip}}}_{\text{positives}}
+ k_2 \!\!\underbrace{\sum_{n\in\mathcal{N}(i)} e^{s_{in}}}_{\text{hard negatives}}
+ k_1 \!\!\underbrace{\sum_{p\in\mathcal{P}(i)} e^{-s_{ip}}}_{\text{hard positives}}.
\end{align}

The per-anchor loss for anchors with at least one positive is
\begin{align}
\ell_i = - \log 
{\displaystyle \sum_{p\in\mathcal{P}(i)} e^{s_{ip}}}
/ {\displaystyle D_i}, \ \text{defined only if }|\mathcal{P}(i)|>0.
\end{align}

Averaging over all valid anchors gives the final TCL objective:
\begin{align}
\mathcal{L}_{\mathrm{TCL}} 
&= \frac{1}{|\mathcal{V}|} \sum_{i\in\mathcal{V}} \ell_i \ \text{with} \  \mathcal{V} = \{\,i\in\mathcal{B}: |\mathcal{P}(i)|>0\,\}
\end{align}

The first term $\sum_{p}\!e^{s_{ip}}$ pulls same-class pairs closer, 
while $k_2\!\sum_{n}\!e^{s_{in}}$ increases repulsion from 
difficult negatives and $k_1\!\sum_{p}\!e^{-s_{ip}}$ penalizes 
poorly aligned positives. 
To prevent instability at early epochs, $k_1$ is linearly ramped over the first $R$ epochs using $k_1(e)=(1-t)k_1^{(0)}+t\,k_1^{(T)}$ with $t=\min\!\big(1,\frac{e-1}{R}\big)$, where $e$ is the current training epoch.
This adaptive contrastive formulation improves intra-class cohesion and inter-class separation in the learned embedding space.

\subsubsection{\textbf{Dictionary-Based Contrastive Learning (DBCL)}}
While TCL operates within a mini-batch, DBCL extends contrastive learning by maintaining a per-class memory bank that stores 
historical embeddings and prototypes, enabling richer and more stable contrastive sampling across batches.
We maintain a per-class memory of fixed-size queues
$\{\mathcal{Q}_c\}_{c\in\mathcal{C}}$ (capacity $K$) and an exponential-momentum prototype
$\boldsymbol{\mu}_c$ for each class $c$ \cite{xu2024attention}. After each batch,
$\boldsymbol{\mu}_c \leftarrow \mathrm{norm}\ \!\big(m\,\boldsymbol{\mu}_c + (1{-}m)\,\bar{\mathbf{z}}_c\big)$,
where $\bar{\mathbf{z}}_c$ is the mean of current $l_2$-normalized embeddings with label $c$ (if any) and $m\in[0,1)$.

For an anchor embedding $\mathbf{z}_i$ with label $y_i$, we retrieve:
(i) the $P$ most similar stored positives
$\mathcal{P}_{\mathrm{mem}}(i)=\mathrm{TopP}\big(\mathcal{Q}_{y_i};~\mathbf{z}_j^\top\mathbf{z}_i\big)$,
(ii) the $Q$ hardest negatives
$\mathcal{N}_{\mathrm{mem}}(i)=\mathrm{TopQ}\big(\!\bigcup_{c\neq y_i}\!\mathcal{Q}_c;~\mathbf{z}_j^\top\mathbf{z}_i\big)$,
and (iii) the class prototype $\boldsymbol{\mu}_{y_i}$. Here “hardest” means largest cosine similarity to the anchor.
Let $s_{ij}=\mathbf{z}_i^\top\mathbf{z}_j/\tau_{\mathrm{dict}}$ with temperature $\tau_{\mathrm{dict}}>0$.
Averaging over anchors with at least one retrieved positive (index set $\mathcal{U}$),
the dictionary loss mirrors TCL and adds a prototype-attraction term (weight $w_p$):
\begin{align}
\mathcal{L}_{\mathrm{dict}}
&= - \frac{1}{|\mathcal{U}|}\sum_{i\in\mathcal{U}}
\log \frac{\mathrm{N}_i}{\mathrm{D}_i},
\\[-2pt]
\mathrm{N}_i
&= \sum_{p\in\mathcal{P}_{\mathrm{mem}}(i)} \exp\!\big(s_{ip}\big)
~+~ w_p\,\exp\!\Big(\tfrac{\mathbf{z}_i^\top \boldsymbol{\mu}_{y_i}}{\tau_{\mathrm{dict}}}\Big),
\\[-2pt]
\mathrm{D}_i
&= \mathrm{N}_i
~+~ \sum_{n\in\mathcal{N}_{\mathrm{mem}}(i)} \exp\!\big(s_{in}\big).
\end{align}
The top-$P$ stored positives from $\mathcal{Q}_{y_i}$ provide \emph{same-class neighbors}
beyond the mini-batch, strengthening attraction to class-consistent features.
The top-$Q$ negatives from $\cup_{c\neq y_i}\mathcal{Q}_c$ inject \emph{hard negatives} (most similar different-class examples), sharpening the decision boundary.
The prototype term $w_p\,\exp(\mathbf{z}_i^\top \boldsymbol{\mu}_{y_i}/\tau_{\mathrm{dict}})$ acts as a stable
class “centroid” pull, particularly helpful when few positives are currently in memory.
\noindent\textit{Blending with TCL.}
The overall contrastive objective is
\[
\mathcal{L}_{\mathrm{con}}
~=~ (1-\alpha_{\mathrm{dict}})\,\mathcal{L}_{\mathrm{TCL}}
~+~ \alpha_{\mathrm{dict}}\,\mathcal{L}_{\mathrm{dict}},
\]
with $\alpha_{\mathrm{dict}}\in[0,1]$ (optionally linearly ramped early in training).

\subsubsection{\textbf{Pre-Pruning Objective}}
The total pre-pruning loss combines classification and contrastive terms:
\begin{align}
\mathcal{L}_{\mathrm{pre}} = (1-\lambda)\,\mathcal{L}_{\mathrm{CE}} + \lambda\,\mathcal{L}_{\mathrm{con}}
\end{align}

where $\lambda\in[0,1]$ balances class discrimination and embedding alignment.
This stage yields a well-trained \emph{teacher model} used later for pruning and knowledge distillation.

\subsection{Post-Pruning Stage}
The post pruning stage comprises of a teacher snapshot, knowledge distillation, fine tuning, and deployement.

\subsubsection{\textbf{Teacher Snapshot and Pruning}} After pre-training, a frozen teacher model \(t(\cdot)\) is created by snapshotting the pre-pruned model \cite{hinton2015distilling}. This teacher model guides the pruned student model during the distillation process, helping the student retain the performance learned before pruning.

\subsubsection{\textbf{Knowledge Distillation (KD) and Fine-Tuning}}
During \emph{post-pruning} step, we add KD, and the model is fine-tuned under the supervision of a frozen teacher model. For the KD view set \(\mathcal{V}_{\mathrm{kd}}\in\{\{1\},\{1,2\}\}\) and temperature \(T>0\), define
\begin{align}
\mathbf{o}_t^{(v)} &= t\!\left(\mathbf{X}^{(v)}\right), &
p_t^{(v)} &= \mathrm{softmax}\!\left(\frac{\mathbf{o}_t^{(v)}}{T}\right), \\
\mathbf{o}_s^{(v)} &= f_{\theta}\!\left(\mathbf{X}^{(v)}\right), &
p_s^{(v)} &= \mathrm{softmax}\!\left(\frac{\mathbf{o}_s^{(v)}}{T}\right).
\end{align}
The KD loss averages the KL divergence across KD views and uses the standard \(T^{2}\) factor:
\begin{align}
\mathcal{L}_{\mathrm{KD}}
&= \frac{T^{2}}{|\mathcal{V}_{\mathrm{kd}}|}\!
\sum_{v\in\mathcal{V}_{\mathrm{kd}}}
\mathrm{KL}\!\big(p_t^{(v)} \,\|\, p_s^{(v)}\big).
\end{align}

The classification objective \emph{used in post-pruning fine-tuning} blends label-smoothed CE and KD:
\begin{align}
\mathcal{L}_{\mathrm{cls}}
&= (1-{\mathrm{kd_\alpha}})\,\mathcal{L}_{\mathrm{CE}}
\;+\; {\mathrm{kd_\alpha}}\,\mathcal{L}_{\mathrm{KD}},
\end{align}
where \({\mathrm{kd_\alpha}}\in[0,1]\) controls the trade-off. The final objective function for post-pruning fine-tuning is 
\[
\mathcal{L}_{\mathrm{post}}
= (1-\lambda)\,\mathcal{L}_{\mathrm{cls}}
\;+\; \lambda\,\mathcal{L}_{\mathrm{con}}
\]

\subsubsection{\textbf{Deployment}}

After pruning, KD and fine-tuning, the projection head is removed and only the backbone $f_\theta$ and classifier are retained. The optimized model is then ready for real-time deployment on edge devices for GNSS jamming classification, ensuring both high performance and low computational overhead.


\begin{table}[t]
\centering
\captionsetup{font=small}  
\caption{Model size, Memory Footprint and Inference time}
\label{tab:model_static}
\small
\setlength{\tabcolsep}{6pt} 
\begin{tabularx}{\columnwidth}{@{}l
  S[table-format=2.1]
  S[table-format=2.2]
  S[table-format=1.2]@{}}
\toprule
\text{Model} & \multicolumn{1}{c}{\textbf{Params}} & \multicolumn{1}{c}{\textbf{Memory}} & \multicolumn{1}{c}{\textbf{Inference Time}} \\
 & \text{(mil.)} & \text{(MB)} & \text{(ms/sample)} \\
\midrule
Hybrid CNN          & 3.2  & 12.39 & 0.43 \\
MobileViT          & 2.4  &  9.54 & 0.19 \\
ResNet-18            & 11.1 & 42.73 & 0.22 \\
DBCL(Pre-Pru.)     & 11.5 & 43.98 & 0.22 \\
DBCL(Post-Pru.)    & \textbf{2.3}  &  \textbf{9.2} & \textbf{0.12} \\
\bottomrule
\end{tabularx}
\end{table}


\begin{table}
\centering
\captionsetup{font=small}  
\caption{Accuracy (\%) vs. Training samples per class}
\label{tab:acc_grid}
\renewcommand{\arraystretch}{1.5}  
\large
\resizebox{\columnwidth}{!}{%
\begin{tabular}{c S S S S S}
\toprule
\textbf{Samples/class} & \textbf{Hybrid CNN} & \textbf{MobileViT} & \textbf{ResNet-18} & \textbf{DBCL Pre-Prun} & \textbf{DBCL Post-Prun.} \\
\midrule
50   & 69.5 & 55.8  & 85.8 & 91.2 & 89.1 \\
300  & 96.1 & 92.1  & 97.1 & 98.2 & 97.5 \\
1000 & 98.6 & 98.0  & 98.7 & 98.8 & 98.6 \\
1500 & 98.7 & 98.4  & 98.8 & 99.2 & 98.9 \\
2000 & 98.8 & 98.1  & 99.1 & 99.1 & 99.0 \\
3000 & 98.8 & 98.8  & 98.8 & 99.2 & 99.1 \\
\bottomrule
\end{tabular}%
    }
\end{table}

\begin{figure}
    \centering
\includegraphics[width=\columnwidth]{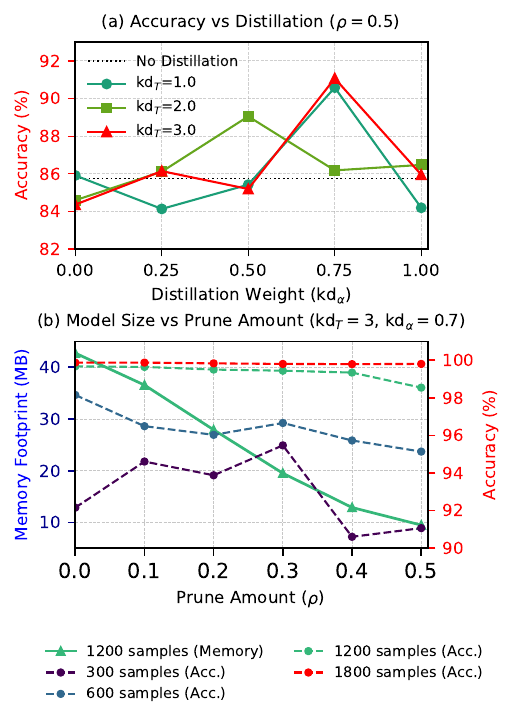}
    \caption{(a) Accuracy variation with KD at prune ratio 0.5 with 1000 samples per class
    (b) Model footprint and accuracy variation with pruning amount for fixed KD}
    \label{fig:kd_prune_subfigs}
\end{figure}

\begin{figure*}[t]
\centering
\captionsetup[subfigure]{justification=centering, labelformat=parens, labelsep=space}
\subfloat[ResNet-18]{%
    \includegraphics[width=0.265\textwidth]{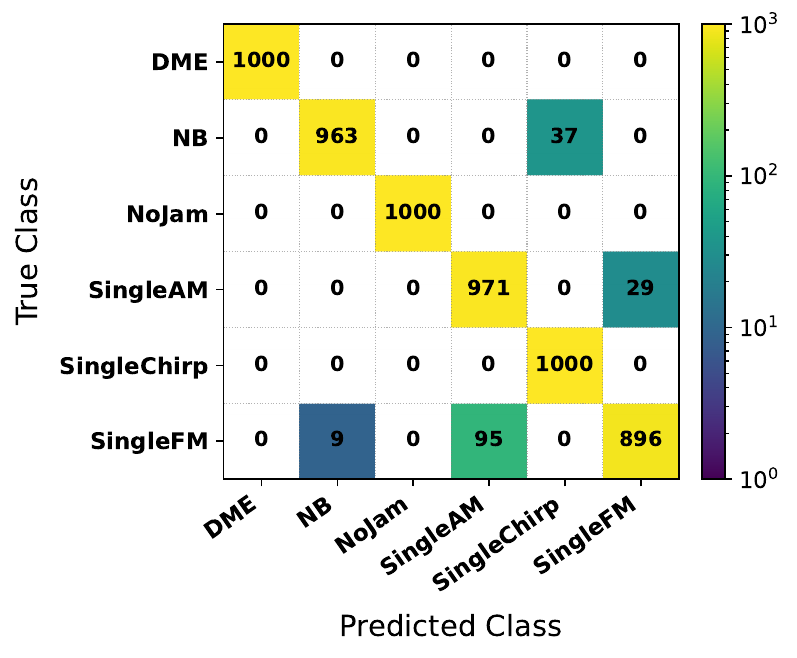}}
\hfill
\subfloat[MobileViT]{%
    \includegraphics[width=0.265\textwidth]{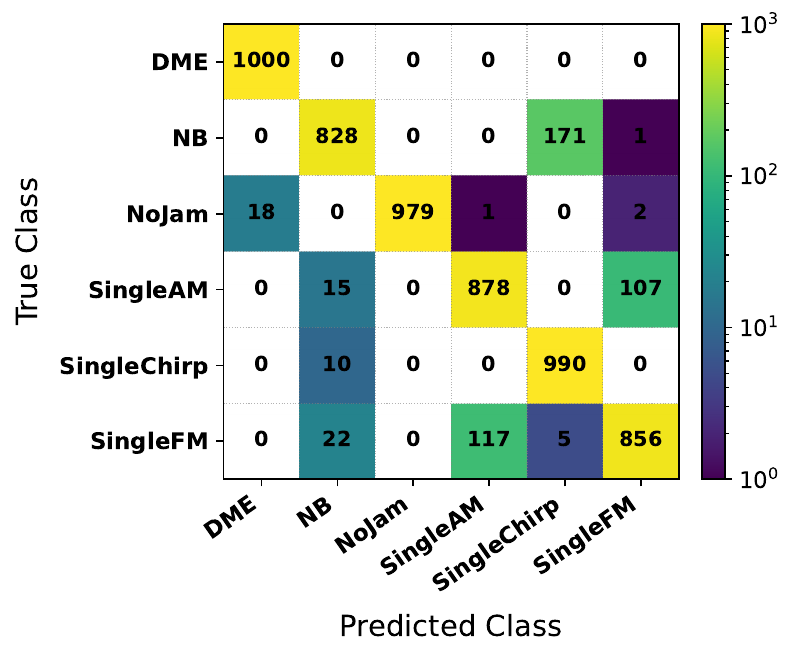}}
\hfill
\subfloat[Hybrid CNN]{%
    \includegraphics[width=0.265\textwidth]{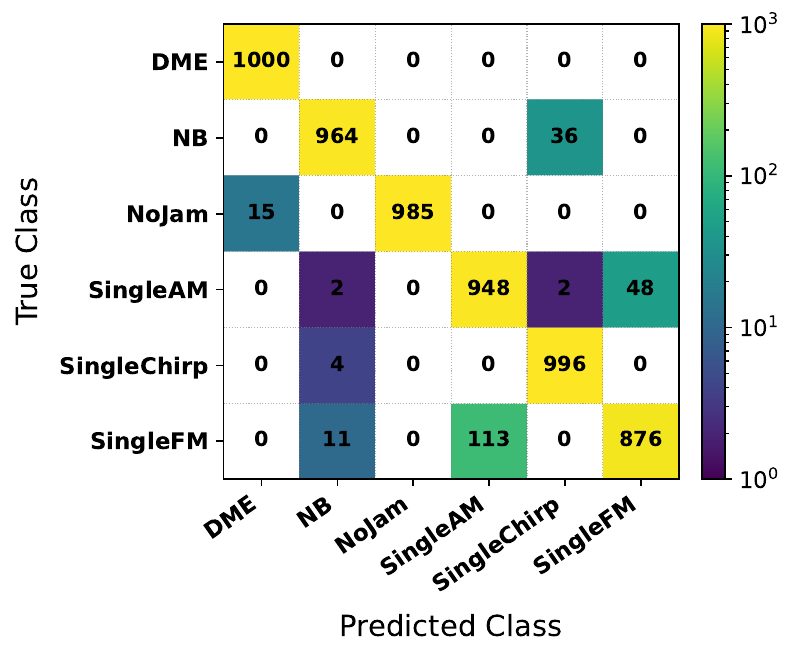}}

\vspace{0.5em}

\subfloat[DBCL-Before Pruning]{%
    \includegraphics[width=0.265\textwidth]{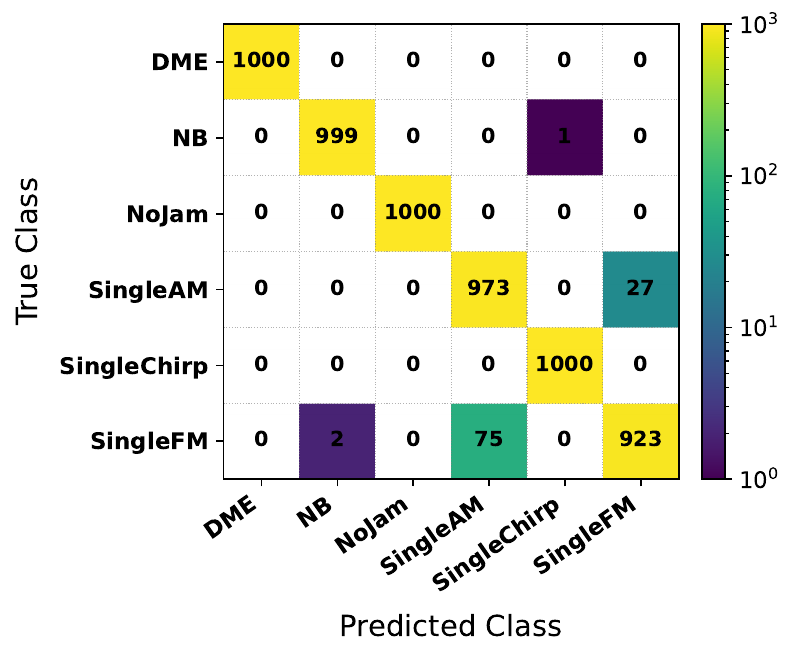}}
\hspace{0.07\textwidth}
\subfloat[DBCL-After Pruning]{%
    \includegraphics[width=0.265\textwidth]{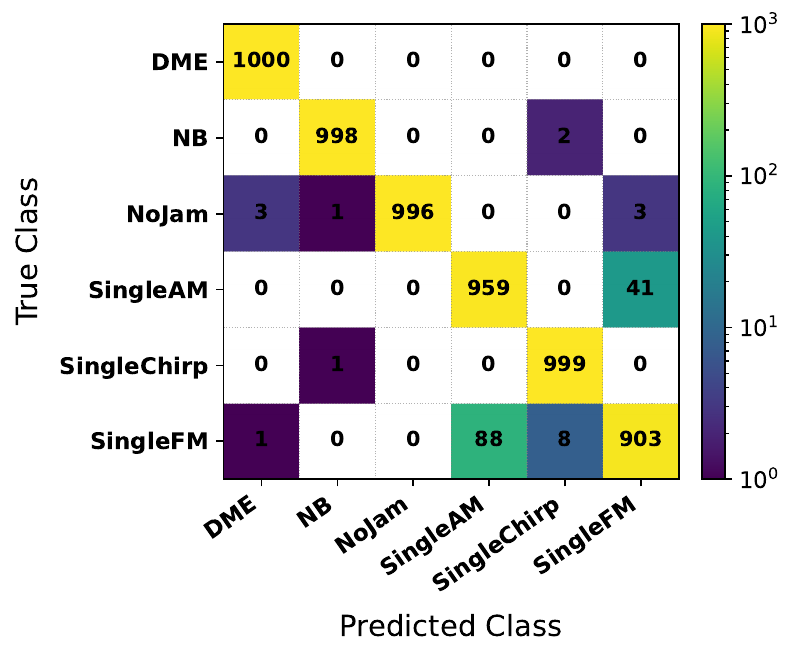}}

\caption{Confusion matrices of different models evaluated on the GNSS interference dataset with 300 samples per class.}
\label{fig:confusion_comparison}
\end{figure*}





\section{Simulation Results}
All simulations were carried out using the PyTorch 2.7.1 (+cu126) deep learning framework on a workstation equipped with an NVIDIA GeForce RTX 2080 Ti GPU (16 GB RAM). The results of our simulations are summarized in Table~\ref{tab:model_static}, which compares the model size and computational efficiency of the proposed DBCL framework with baseline architectures. Notably, the post-pruned DBCL achieves a substantial reduction in both parameters (from 11.5 M to 2.3 M) and memory footprint (from 43.9 MB to 9.2 MB), while delivering the fastest inference speed of 0.12 ms/sample. This 77\% reduction in memory footprint highlights DBCL's suitability for resource-constrained edge devices without sacrificing computational performance.


The results in Table~\ref{tab:acc_grid} present the classification accuracy of the models across different training sample sizes per class, evaluated on 6000 test samples with 1000 samples per class. The DBCL consistently outperforms hybrid CNN, MobileViT, and ResNet-18 models, especially under low-data sample conditions. At 50 samples/class, it achieves 91.2\% (pre-pruning) and 89.1 \% (post-pruning), demonstrating strong generalization even with limited data. As sample size increases beyond 1000 per class, all models converge near 99\% accuracy. However, DBCL maintains the best trade-off between accuracy, compactness, and inference speed, proving its effectiveness for real-time GNSS jamming detection on embedded edge platforms.

We examined the combined effect of pruning and knowledge distillation on accuracy and model compactness. With a fixed pruning ratio of 0.5, varying the KD parameters ($\mathrm{kd}_{\alpha}=0.75$, $\mathrm{kd}_T=3$) achieved the highest accuracy, while the memory footprint remained constant as shown in Fig.~\ref{fig:kd_prune_subfigs}(a). In contrast, adjusting the pruning ratio reduced the model size from over 40 MB to around 10 MB, maintaining accuracy above 90\% across different data samples per class as shown in Fig.~\ref{fig:kd_prune_subfigs}(b). These results validates that the proposed pruning–distillation strategy achieves an excellent trade-off between compactness and performance, 
enabling efficient GNSS jamming classification on embedded edge platforms.

The confusion matrices in Fig.~\ref{fig:confusion_comparison} illustrate the classification performance of different models on the GNSS interference dataset with 300 training samples per class. Compared to baseline methods such as Hybrid CNN, MobileViT, and ResNet-18, the proposed DBCL framework demonstrates superior class discrimination and minimal misclassification across all interference types. Both pre-pruning and post-pruning variants of DBCL show highly consistent results, accurately identifying all interference classes, including challenging categories like SingleChirp and SingleAM, which exhibit higher confusion in other models. These results confirm that DBCL effectively learns more separable and robust feature embeddings, achieving high accuracy while maintaining compactness suitable for real-time, edge-based GNSS jamming detection.



\section{Conclusion}

In this work, we presented a Dictionary-Based Contrastive Learning (DBCL) framework for GNSS jamming detection that unifies contrastive embedding, transfer learning, and model compression to achieve high accuracy with lightweight deployment. By integrating tuned contrastive and dictionary-based objectives with structured pruning and knowledge distillation, the proposed method achieves robust representation learning and significant reductions in model size and inference latency. Simulation results demonstrated that DBCL outperforms Hybrid CNN, MobileViT, and ResNet-18 baselines, particularly in low-data conditions, while maintaining strong generalization at scale. The framework’s compactness and efficiency confirm its suitability for real-time, energy-constrained edge applications, paving the way for scalable and intelligent GNSS interference detection in next-generation navigation systems.

\newpage
\makeatletter
\patchcmd{\thebibliography}{#1}{99}{}{}
\makeatother

\IEEEtriggeratref{10}
\IEEEtriggercmd{\RightColNudge}
\bibliographystyle{IEEEtran}  


\bibliography{reference}     

\end{document}